# WLAN location system:

## Background theories and future directions

Debabala Swain, CIT        S.K.Routray, KIST        R.R.Mohanty, KIST

S.P.Panigrahi, KEC        P.K.Dash, KIST        S.K.Dash, KIST

Bhubaneswar, Orissa, India, siba_panigrahy15@rediffmail.com

*Abstract*—This paper presents background theories and required steps towards preparation of a WLAN location system. This paper targets on a software project and intention behind this paper is to motivate the young researchers in the area.

## I. INTRODUCTION

There are two essential background theories relevant our design using empirical model. One is how to enable the project hardware to retrieve information from surrounding stations. The other is how to implement k-nearest neighbor algorithm into the system. This paper covers these two issues.

## II. BACKGROUND THEORIES

### A. Wireless LAN Scanning

In the Wireless LAN, there are two ways to scan and retrieve the signal information from surrounding stations and access points: RF Monitoring and Active Probing.

#### 1) RF Monitoring

The RF monitoring (RFMON) technique is to make uses of the beacon frames sent out periodically from the 802.11 stations. A Wireless LAN card with an appropriate driver support can be used to continuously capture the beacon frames in the air and the frames then can be passed to any packet analyser that recognises the beacon frame formats, to retrieve the information in the beacon frames. In most cases, the beacon frame contains signal information related the frame transmitter and the receiver.

While a Wireless LAN card is in RFMON mode, it will not be able to transmit any frames but only listen in the air medium to capture traffic. This limits the client to reporting only current or recorded network traffic. This can be a problem in a Wireless LAN position system design that wants to support client and server mode. The signal information received by the client will be immediately transmitted to the central server.

In Linux, a Wireless LAN card uses the RFMON via support of its device driver and Linux Wireless Extensions (later discussed). As to the packet analyzers, there are three well-known packet-sniffing applications that can run in Linux systems and will support RF monitoring and display the information in an 802.11 beacon frame after decoding. They are nets umber [1], ethereal [2] and prisms tumbler [3]. In particular, prism-stumbler is the application developed directly to support the Linux Familiar embedded operation system.

#### 2) Active Probing

The active probing method uses IEEE 802.11 specific "probe request frames" on each channel where it is able to detect wireless activity. When an access point comes within range of a client station and receives a probe request frame the access point will typically have to respond with a probe response frame. The probe response frame will contain various information similar to the beacon frame. The active probing method for network discovery is the easiest to implements. Unlike the RF monitoring, the device that uses the active probing method is able to transmit and receive data at the same time.

However, this method has a disadvantage compared to RF monitoring. It is unable to discover wireless networks that are configured not to advertise their existence (Not transmitting any probe response frame) via a cloaked ESSID (Extended Service Set Identification) configuration. In the Wireless LAN positioning design, that implies all the reference access points have to be in broadcast, which sometimes might not be desired due to security measurements.

Similar to the RF monitoring, the active probing has to be supported by the Wireless LAN card's device driver, via the Linux Wireless Extensions. With the active probing wireless extension, the device can easily retrieve station's signal information by using Linux Wireless Tools. There is no need of a packet analyser.

#### 3) Wireless Extensions and Wireless Tools for Linux

The Linux Wireless Extensions [4] and the Linux Wireless Tools [5] are an Open Source project sponsored by Hewlett Packard since 1996, and build with the contribution of many Linux users all over the world.

The Wireless Extensions is a generic application program interface (API) allowing a driver to expose to the user space configuration and statistics specific to common Wireless LANs. The beauty of it is that a single set of tool can support all the variations of Wireless LANs, regardless of their type (as long as the driver supports Wireless Extensions). One more advantage is these parameters may be changed on the fly without restarting the driver (or Linux).

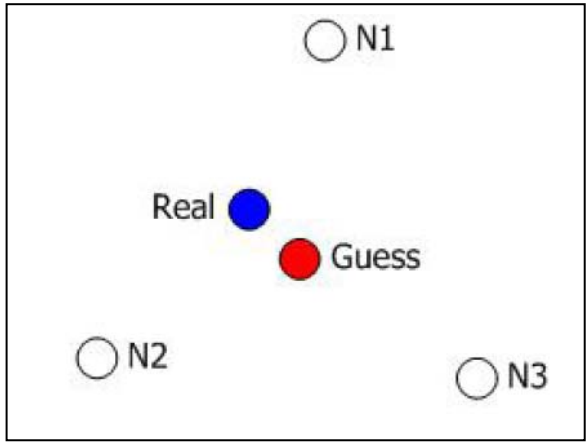

Figure 1. K-Nearest neibor Search (k=3).

The Wireless Tools is a set of tools allowing manipulating the Wireless Extensions. They use a textual interface and are rather crude, but aim to support the full Wireless Extensions. The latest version of Wireless Tools (Version 26) contains four commands:
- iwconfig – used to manipulate the basic wireless parameters.
- iwlist (formerly part of iwspy) – used to allow to list addresses, frequencies, bit-rates and other signal related information
- iwspy – used to allow to get per node link quality
- iwpriv – used to allow to manipulate the Wireless Extensions specific to a driver

In recent versions of Wireless Tools, iwlist supports buffering AP information of up to 64 entries, in the project's design it implies the device can detect maximum 64 access points at once (compared to 8 in the Amulet project). Jean Tourrilhes wrote both Linux Wireless Extensions and Wireless Tools.

*B. K-Nearest Neibor Algorithm*

K-Nearest Neighbour is a simple algorithm that stores all available examples and searches for a supplied entry the $k$ number of the examples that have the highest similarity measures. The examples and the supplied entry are numerical values in the same formats and the "vector distance" between the supplied entry and each example determines the similarity measure. The vector distance is defined by Euclidian distance. If the example vector is said to be $(X_e, Y_e, Z_e)$ and the supplied data entry is $(X_s, Y_s, Z_s)$ then the vector distance between the two vectors is $\sqrt{(X_e - X_s)^2 + (Y_e - Y_s)^2 + (Z_e - Z_s)^2}$; the lower the vector distance is, the higher similarity measure is between this example and the supplied entry.

When using in the Wireless LAN positioning, a database may contain a set of examples where an example is a series of signal strength values from a fixed number of access points. The examples are collected during the radio map construction and each example corresponds to a known X-Y coordinates on the map. A client scans the signal values from the same access points and stores into an entry in the same order. This entry is then compared with the examples in the database and retrieves a k number of nearest examples, using the K-Nearest Neighbor algorithm. The locations corresponding to these nearest examples should then be the ones closest to the location of the client. Of course, when k = 1, the location of the nearest neighbor can be said to be the location of the client. However, since the radio map is usually made such that a fixed point is always a constant distance from the other fixed point, the real location of nearest neighbor may not be at a fixed point but among various fixed points. Therefore normally it is more accurate to decide the location of the client by averaging the locations of k fixed points, where k is kept small. The figure next page is an illustration of how averaging multiple nearest neighbors (N1, N2, N3) can lead to a guess point that is closer to the true location than any of the neighbors is individually. However, for large k, accuracy degrades rapidly because points far removed from the true location also are included in the averaging procedure, thereby corrupting the estimate.

Fortunately, there is no need to start from the scratch in order to implement the K-nearest neighbor algorithm in the system design. ANN (Approximate Nearest Neighbor) [6] is library written in C++, which supports data structures and algorithms for both exact and approximate nearest neighbor searching in arbitrarily high dimensions. ANN was successfully compiled and run in the embedded Linux OS.

### III. PREPARING FOR A PROJECT

For any software-based project, the preparation stage is most likely about setting up a working programming environment. And normally when an application is to be developed for embedded type of devices, the programming environment will on a remote desktop PC as it provides large storage space (to store cross-compiler and program source codes) and faster CPU. The application would be written on the PC, and cross-compiled using the specific cross-compiler for the target platform and directly sent to the target device for executions.

To design the application of this project, a cross-compiling environment was required to be setup. In this section, all the components in order to setup the environment are discussed.

*A. iPAQ H3630 with Orinoco-based Wireless LAN PC Card*

Our target device is an iPAQ H3630 [7]. The handheld equips a 32-bit Intel StrongARM SA-1110 microprocessor, running at 206 MHz on a 100 MHz memory bus. It contains 32MB RAM and 16MB ROM to store the embedded operating system and applications. To make the H3630 Wireless LAN enable, a compatible PCMCIA expansion pack was added to the handheld to use Orinoco Silver Wireless LAN PCMCIA card [8]. The Orinoco Silver is an 802.11b standard WLAN client supported by various Linux drivers. Serial and USB cradles for H3630 were both used throughout the project development to transfer data from the regular desktop PCs to the handheld.

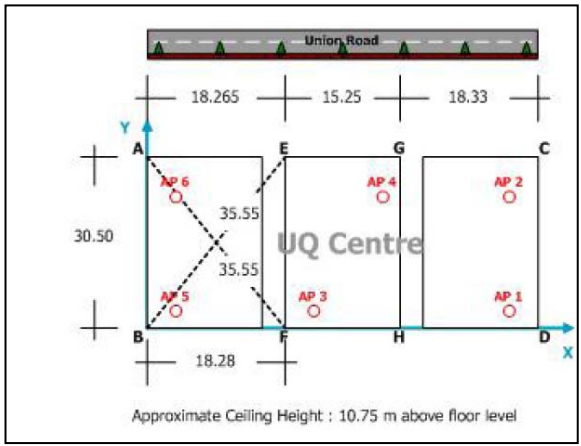

Figure 2.  UQ-Centre location map

### B. Familiar and OPIE

Due to the project needs, the operating system in the iPAQ H3630 was replaced with Familiar (Linux-based embedded operating system) from the original Microsoft Windows CE. Installing atop the Familiar is a graphical user interface environment called Open Palmtop Integrated Environment (OPIE). In order to run the Familiar OS, the original boot-loader on the handheld was also replaced. The OS replacement can be made by following the instructions on Handhelds.org website.

Handhelds.org How-tos –
http://www.handhelds.org/minihowto/index.html

Throughout the development, various versions of familiar were installed and tested, mainly trying to find a Wireless LAN driver for the Orinoco Silver card that could support either RF monitoring or active probing. It has taken a long period of time learning how to recompile the embedded Linux kernel of the Familiar and the device drivers. The version 7.1 was the version was used.

Familiar Official Website – http://familiar.handhelds.org/

Comparing to the installing experience with the familiar, developing a graphical user interface under OPIE has been a smooth process. OPIE is a 'fork' of Qtopia environment developed by Trolltech [9]. It is a completely Open Source based user graphical environment for PDAs and other devices running Linux. The OPIE website provides a complete documentation of the OPIE application programming interfaces (APIs), including the comprehensive start-up tutorials.

In addition, through the help of the people on both the OPIE mailing list and IRC channel, most of the GUI related development questions were solved without much struggling.

OPIE Official Website – http://opie.handhelds.org/

### C. Remote PC and Cross-Compiler

The PC used to write and compile the application is a Compaq Presario 1700 series laptop. The laptop was running a Pentium III 1 GHz mobile microprocessor, with 512 MB of RAM. Red Hat Linux 9 was installed and used as the cross-compiling development platform. With this setup, there was no noticeable performance issue with compiling the C and C++ codes for the project.

The search of a cross-compiler that would work on the development platform has caused nothing but headaches. There were various versions of cross-compilers on the handhelds.org (the official website that hosts all the Linux embedded developments) FTP server and none of them worked. Eventually, through the OPIE SDK Cookbook website, a RPM (Redhat Package Manager) based cross-compiler was found that worked beautifully on the project's setup.

OPIE SDK Cookbook –
http://www.zaurus.com/dev/tools/other.htm
Cross-compiler for StrongARM –
http://www.zaurus.com/dev/tools/other.htm

### D. Wireless LAN Linux Drivers

There were two Linux drivers found that worked for the Orinoco Silver Wireless LAN card and supported the RF monitoring and active probing wireless extensions. And using in conjunction with Linux Wireless Tools and Linux packet analysers, both drivers were able to provide the desired functionality to retrieve signal information from the access points.

Strong ARM patched rinoco_cs driver supported the RF monitoring. The driver was found installed in all the recent versions of the Familiar distributions. However, later testing showed that only the drivers in Familiar version 7.0 and newer provided the stable RF monitoring (did not crash the operating system).

The presented orinoco_cs drivers in the Familiar distributions (through ipkgfind)
http://ipkgfind.handhelds.org/result.phtml?query=orinoco&searchtype=package§ion=

mwvlan_cs driver used the support of active probing. The mwvlan_cs driver obtained from the Mwvlan website was manually patched and compiled with the Familiar source. In a note, the Familiar FTP provided the StrongARM version of the driver in binary format; however, the binary has never worked on project's hardware.

Mwlvan_cs driver –
http://www.cs.umd.edu/~moustafa/mwvlan/mwvlan.html.

### E. UQ Centre Hall

The part of preparation involved setting up the testing environment – UQ Centre Hall. There were a total of 6 802.11b access points installed on the top of the inner roof and the roof was made of thick metal, which incurred strong pathloss to the transmitting radio signals.

Works presented by Dr. Gerd R Dowideit, the dimension of the area under test was known to be 30.5 by 52 metres. The positions (in x and y) of the access points were also roughly located as shown on the 2-D map below. Each access point that has been uniquely numbered was about 10.75 metres (z axis) above the floor level.

## IV. CONCLUSION

This paper presented the background theories and possible directions for preparing a WLAN location system.